\documentclass[12pt]{article}

\usepackage{latexsym}
\usepackage{amssymb,amsfonts,amsmath}
\usepackage{graphicx} 
\usepackage{indentfirst}
\usepackage{bbm}
\usepackage{amssymb}
\usepackage{verbatim}
\usepackage{amsmath, amsthm,amssymb}
\usepackage{mathrsfs}
\usepackage{hyperref}
\usepackage{amsfonts}
\usepackage{dsfont}
\usepackage{cite}
\usepackage{xcolor}
\usepackage[multiple]{footmisc}

\parskip=\medskipamount

\newcommand {\cD}{{\cal D}}
\newcommand {\cE}{{\cal E}}

\newcommand {\cL}{{\cal L}}
\newcommand {\cM}{{\cal M}}
\newcommand {\cN}{{\cal N}}

\newcommand {\cP}{{\cal P}}

\def\a{\alpha}

\def\b{\beta}
\def\c{\chi}
\def\d{\delta}

\def\f{\phi}
\def\g{\gamma}
\def\G{\Gamma}

\def\j{\psi}
\def\k{\kappa}
\def\l{\lambda}
\def\m{\mu}

\def\o{\omega}
\def\p{\pi}
\def\q{\theta}
\def\r{\rho}
\def\s{\sigma}

\def\x{\xi}
\def\z{\zeta}
\def\D{\Delta}
\def\F{\Phi}
\def\J{\Psi}

\def\S{\Sigma}
\def\U{\Upsilon}

\def\tr{{\rm tr}}
\def\rd{{\rm d}}
\def\ri{{\rm i}}
\def\re{{\rm e}}

\newcommand{\ad}{{\dot{\alpha}}}                           
\newcommand{\bd}{{\dot{\beta}}}                            
\newcommand{\ve}{\varepsilon}                            

\newcommand{\pa}{\partial}                           
\newcommand{\hf}{\frac12}

\newcommand{\vf}{\varphi}

\newcommand{\be}{\begin{equation}}
\newcommand{\ee}{\end{equation}}
\newcommand{\bea}{\begin{eqnarray}}
\newcommand{\eea}{\end{eqnarray}}
\newcommand{\non}{\nonumber}
\def\double #1{#1{\hbox{\kern-2pt $#1$}}}

\newcommand{\qb}{{\bar{\theta}}}

\newif\ifdtup

\newcommand{\bsubeq}{\begin{subequations}}
\newcommand{\esubeq}{\end{subequations}}

\numberwithin{equation}{section}

\begin{document}

\begin{titlepage}
\begin{flushright}
September, 2020 \\
\end{flushright}
\vspace{5mm}

\begin{center}
{\Large \bf 
Effective actions for dual massive (super) $p$-forms}
\end{center}

\begin{center}

{\bf Sergei M. Kuzenko and Kai Turner} \\
\vspace{5mm}

\footnotesize{
{\it Department of Physics M013, The University of Western Australia\\
35 Stirling Highway, Perth W.A. 6009, Australia}}
~\\
\vspace{2mm}
Email: \texttt{ 
sergei.kuzenko@uwa.edu.au, 22226913@student.uwa.edu.au}\\

\end{center}

\begin{abstract}
\baselineskip=14pt
In $d$ dimensions, the model for a massless $p$-form in curved space is known 
to be a reducible gauge theory for $p>1$, and therefore its covariant quantisation cannot be carried out using the standard Faddeev-Popov scheme. However, adding a mass term and also introducing a  Stueckelberg reformulation of the resulting $p$-form model, one ends up with an irreducible gauge theory which can be quantised \`a la Faddeev and Popov. We derive a compact expression for the massive $p$-form effective action, $\Gamma^{(m)}_p$,  
in terms of the functional determinants of  Hodge-de Rham operators. 
We then show that the effective actions
$\Gamma^{(m)}_p$ and $\Gamma^{(m)}_{d-p-1}$  differ by a topological invariant. 
This is a generalisation of the known result in the massless case that the effective actions 
$\Gamma_p$ and $\Gamma_{d-p-2}$  coincide modulo a topological term. 
Finally, our analysis is extended to the case of massive super $p$-forms coupled to background ${\cal N}=1$ supergravity in four dimensions. Specifically, we study the quantum dynamics of the following massive super $p$-forms: (i) vector multiplet; (ii)  tensor multiplet; 
and (iii) three-form multiplet. It is demonstrated that the effective actions of the massive vector and tensor multiplets coincide. The effective action of the massive three-form is 
shown to be a sum of those corresponding to  two massive scalar multiplets, modulo a topological term.  
\end{abstract}
\vspace{5mm}

\begin{center}
{\it Dedicated to the memory of Professor Omar Foda}
\end{center}

\vfill

\vfill
\end{titlepage}

\newpage
\renewcommand{\thefootnote}{\arabic{footnote}}
\setcounter{footnote}{0}

\tableofcontents{}
\vspace{1cm}
\bigskip\hrule

\allowdisplaybreaks


\section{Introduction}

The model for a massless gauge two-form in four dimensions was introduced in the mid-1960s by Ogievetsky and Polubarinov  \cite{OP} 
who showed that it describes a spin-zero particle. 
Unfortunately, their work remained largely unknown for a decade. 
 The same model was rediscovered, and generalised, twice in 1974  in the context of dual resonance models \cite{KR,CS}.  However, active studies of gauge $p$-forms in diverse dimensions began only in the late 1970s when it was recognised 
 that such fields naturally occur in supergravity theories,
 see, e.g., \cite{GSO,CSF,CJS} for early publications 
 and \cite{VanN,SalamSezgin,Tanii} for reviews. Gauge $p$-forms are also of special interest in  string theory where they appear in the low-energy effective actions  see, e.g., \cite{GSW,Polchinski,BBS,BLT} for reviews. 
 
 There are two important themes in modern quantum field theory that originated by studying the quantum dynamics of massless gauge $p$-forms: (i) reducible gauge theories; 
 and (ii) quantum equivalence of dual theories. It is appropriate here to briefly recall these developments. 

 For $p>1$, all massless $p$-form actions
 are examples of the so-called reducible gauge theories 
 (following the terminology of the Batalin-Vilkovisky formalism \cite{BV}).
 In the framework of covariant Lagrangian quantisation,
  reducibility  means that the generators of gauge 
 transformations are linearly dependent. 
 This fact has a number of non-trivial implications, which are: (i) gauge-fixing functions are constrained; (ii) ghosts for ghosts are required; and (iii) a naive application of the Faddeev-Popov quantisation scheme leads to incorrect results. Several consistent quantisation procedures have been developed to quantise reducible Abelian  gauge theories such as gauge $p$-forms \cite{Schwarz1,Schwarz2,Siegel,Obukhov,BK88},   
 including the formulations of \cite{Siegel,BK88} which apply in the supersymmetric case.  
 These quantisation schemes are much easier to deal with than the 
  Batalin-Vilkovisky formalism \cite{BV}.\footnote{One of the earliest applications of the 
  Batalin-Vilkovisky formalism \cite{BV} was the Lagrangian quantisation 
\cite{BK87,deAGM} of the Freedman-Townsend model \cite{FreedmanT}.  
Ref. \cite{BK87} was accepted for publication in  Sov. J. Nucl. Phys. in 1987. 
It was subsequently withdrawn shortly before publication, 
after the authors had been informed by a colleague
that the same problem had already been solved elsewhere. Due to a limited access to 
the journals, at the time it was not possible  to verify this information, 
which in fact turned out  to be false. 
}

In $d$ dimensions, two massless field theories describing a  $p$-form and a $(d-p-2)$-form 
are known to be classically equivalent, see, e.g., \cite{Tanii,HL} for reviews.
These theories are dual  in the sense that the corresponding  
actions are related through the use of a first-order (or parent) action, see e.g. \cite{FT}. 
The issue of quantum equivalence of such classically equivalent theories was raised, 
building on the results of \cite{SvN},  in 
1980 by Duff and van Nieuwenhuizen \cite{DvN,Duff}.  They showed, in particular, 
that (i) a massless two-form and a non-conformal scalar in four dimensions give rise to different trace anomalies; and (ii) the corresponding one-loop divergences differ by 
a topological term. These results were interpreted in \cite{DvN} as 
a possible quantum non-equivalence of these dual field realisations. 
The issue was resolved in several publications \cite{Siegel81,FT,GNSZ,BK88} 
in which it was shown that the effective actions of dual massless theories 
in four dimensions differ only by a topological invariant being independent of the spacetime metric. As a result, the dual theories are characterised by the same quantum energy-momentum tensor, $\langle T_{ab} \rangle$, which proves their quantum 
equivalence.\footnote{In the four-dimensional case, the dual two-form and zero-form theories are classically non-conformal. As emphasised in \cite{GNSZ},  
the quantum operator $T^a{}_a$ in such theories
``contains the effects of both classical and quantum breaking and is not equal to the trace anomaly.'' 
In other words, there is no point to compare trace anomalies in classically non-conformal theories.}  
 Analogous results hold in higher dimensions \cite{FT,GKVZ}, as well as for dual 
supersymmetric field theories in four dimensions \cite{GNSZ,BK88} (see also \cite{BK} for a review). It is worth discussing the supersymmetric story in some more detail.

Several important massless $\cN=1$ supermultiplets in four dimensions can be realised in terms of super $p$-forms \cite{Gates} (see also \cite{GGRS}), with the cases $p=0, 2$ and 3 corresponding to the chiral, tensor and three-form multiplets, respectively. The corresponding supersymmetric theories are related either by a  duality transformation or by a superfield reparametrisation. The simplest model for the tensor multiplet \cite{Siegel-tensor} in a supergravity background is given by the action\footnote{Our consideration below can readily be extended to the nonlinear theories which were introduced in \cite{Siegel-tensor} and are obtained by replacement $G^2\to f(G)$. However, such theories are non-renormalisable in general and will not be studied in what follows. It should be pointed out that the duality transformations
for the nonlinear $f(G)$ models were described in \cite{LR83}. 
The special choice of $f(G) \propto G \ln G $
corresponds to the so-called improved (superconformal) tensor multiplet \cite{deWR}.}
\bea
S_{\rm tensor} [\J, \bar \J] = -\hf \int\rd^4x\rd^2\q\rd^2\qb\, E \, \big( G(\J) \big)^2 ~, 
\qquad G(\J) := \hf \big( \cD^\a \J_\a + \bar \cD_\ad \bar \J^\ad \big) ~,
\label{1.1}
\eea
where $\J_\a$ is a covariantly chiral spinor, $\bar \cD_\bd \J_\a =0$. Its dual version
 \cite{Siegel-tensor} 
\bea
S_{\rm chiral} [\F, \bar \F] = \hf \int\rd^4x\rd^2\q\rd^2\qb\, E \, (\F + \bar \F)^2~,
\qquad \bar \cD_\bd \F =0~,
\label{1.2}
\eea
 describes the non-conformal scalar multiplet. 

Let us represent the dynamical variables in \eqref{1.2} as $\F = \cP_+ V$ 
and $\bar \F = \cP_- V$, where $V$ is a real scalar and 
the operators $\cP_+ $ and $ \cP_- $ have the form\footnote{For any scalar superfield $U$,  $\cP_+ U$ is covariantly chiral, and
$\cP_- U$ antichiral.}
\bea
\cP_+ = -\frac{1}{4} \big(\bar \cD^2 - 4R\big) ~, \qquad 
\cP_- = -\frac{1}{4} \big( \cD^2 - 4 \bar R\big) ~.
\label{4.4}
\eea
Then we end up with 
the  three-form multiplet realisation \cite{Gates} of the non-conformal scalar multiplet.   
The corresponding action is 
\bea
S_{\text{3-form}} [V]= \hf \int\rd^4x\rd^2\q\rd^2\qb\, E \, V \big(\cP_+ + \cP_- \big)^2 V ~,
\qquad \bar V =V~.
\label{1.3}
\eea
This theory was studied in \cite{BK88}, see also \cite{BK} for a review.

The models \eqref{1.1} and \eqref{1.2} are dually equivalent \cite{Siegel-tensor}. 
Their quantum equivalence was established in \cite{GNSZ} in the case of  an on-shell 
supergravity background, and in \cite{BK88} for an arbitrary supergravity background. 

Since the three-form multiplet action \eqref{1.3} is obtained from \eqref{1.2} 
by setting $\F  = \cP_+ V$, the physical fields  can be chosen to coincide in both models. 
The main difference between the models \eqref{1.2} and \eqref{1.3} at the component level is that  one of the two real auxiliary scalars in \eqref{1.2} is replaced by (the Hodge dual of)
 the field strength of a three-form in the case of \eqref{1.3}. Being non-dynamical, 
 the three-form is known to generate a positive contribution to the cosmological constant \cite{DvN,OS,ANT,Duff89,DJ,BP}.
 In order to achieve a better understanding of the three-form multiplet model \eqref{1.3}, 
 we describe its dual version. It is obtained
by  starting with the first-order action \cite{BK88}
\bea
S[V,L] = - \int\rd^4x\rd^2\q\rd^2\qb\, E \, \Big\{ \hf L^2 
- L \big(\cP_+ + \cP_- \big) V \Big\} ~,
\label{1.4}
\eea
where $V$ and $L$ are unconstrained real scalars. 
Varying $S[V,L]$ with respect to $L$ leads to the three-form multiplet action \eqref{1.3}. 
On the other hand, varying $V$ gives 
\bea
\big(\cP_+ + \cP_- \big) L =0 \quad \implies \quad
 \cP_+  L = \ri \m~, \qquad 
\m = \bar \m = {\rm const}~. 
\label{1.5}
\eea
This constraint defines a deformed tensor multiplet, in accordance with the terminology of  \cite{K17}. The dynamics of this multiplet is described by the action
\bea
S[{\mathfrak L}] = - \hf \int\rd^4x\rd^2\q\rd^2\qb\, E \, {\mathfrak L}^2 ~, 
\qquad  \cP_+  {\mathfrak L} = \ri \m~, \qquad 
\m = \bar \m = {\rm const}~. 
\label{1.6}
\eea
At the component level, the main manifestation of the deformation 
parameter $\m$ in \eqref{1.6} is the emergence of a positive cosmological constant. 
Unlike \eqref{1.6}, no parameter $\m$ is present in the action \eqref{1.3}. However, 
$\m$ gets generated dynamically, since the general solution of 
the equation of motion for \eqref{1.3} contains such a parameter, 
\bea
\big(\cP_+ + \cP_- \big)^2 V  =0 \quad \implies \quad 
\cP_+ \big(\cP_+ + \cP_- \big) V = \ri \m~,
\eea
 with $\m$ a real parameter. On the mass shell, we can identify 
 $\big(\cP_+ + \cP_- \big) V =\mathfrak L$.
The effective actions corresponding to different values of $\m$ differ by a cosmological term. The authors of \cite{BK88} made use of the choice $\m=0$ and demonstrated 
that the effective actions  $\G_{\text{chiral}}$ and  $ \G_{\text{3-form}}$, which 
correspond to the locally supersymmetric models \eqref{1.2} and \eqref{1.3}, 
differ by a topological invariant. 

It should be pointed out that general duality transformations with three-form 
multiplets and their applications were studied in \cite{GLS,FLMS,BFLMS}.

So far we have discussed the models for massless $p$-forms and their supersymmetric extensions. Massive antisymmetric tensor fields were discussed in the physics literature 
even earlier than the massless ones.
Kemmer in 1960 \cite{Kemmer}, 
and independently Takahashi and Palmer in 1970 \cite{TP}, 
showed  that the massive spin-1 particle can be described using a 2-form field.  
Further publications on massive antisymmetric fields 
\cite{CS,DW,CF,Curtright,Townsend,Foda,CFG} revealed, in particular, 
that a massive $p$-form in $d$ dimensions is dual 
to a massive $(d-p-1)$-form.\footnote{Massive $p$-forms naturally occur in the framework of string compactifications with non-trivial background fluxes
\cite{LM,GL}.}
This raised the issue of quantum equivalence of dual models.  
Some quantum aspects of massive $p$-forms  were studied 
using the worldline approach in \cite{BBG1,BBG2}.
In the important work by Buchbinder, Kirillova and Pletnev \cite{BKP}, 
the quantum equivalence of classically equivalent massive $p$-forms in four dimensions
was established. In the present work we extend the results of \cite{BKP} to $d$ dimensions. 
Our proof of the quantum equivalence of dual theories in $d=4$ differs from the one given in 
\cite{BKP}. Our approach is also extended to the case of massive super $p$-forms coupled to background ${\cal N}=1$ supergravity in four dimensions. Specifically, we study the quantum dynamics of the following massive super $p$-forms: (i) vector multiplet; (ii)  tensor multiplet; 
and (iii) three-form multiplet. In particular, we demonstrate that the effective actions of the massive vector and tensor multiplets coincide. 

Massive super $p$-forms have recently found numerous applications, 
including the effective description of gaugino condensation
\cite{BDQQ,FGS,CKL,BLS}, inflationary cosmology \cite{Dudas}, and effective field theories from string flux compactifications \cite{LMMS} (see also \cite{Lanza} for a review).
Here we do not attempt to give a complete list of works on massive super $p$-forms and their applications. However it is worth mentioning those publications in which such supermultiplets were introduced in the case of four dimensional $\cN=1$ supersymmetry.
Massive tensor and vector multiplets coupled to supergravity were studied in  \cite{Siegel-tensor,VanProeyen,CFG}. Tensor multiplets with complex masses were studied in  \cite{DF,LS,K-tensor}. To the best of our knowledge, 
 a massive three-form multiplet was first discussed in \cite{BK}, although a massive three-form is contained at the component level in one of the models introduced by Gates and Siegel \cite{GS}.

This paper is organised as follows. In Section 2 we derive 
 effective actions   $\G^{(m)}_p $ for massive $p$-form models 
in $d$-dimensional curved spacetime.  
We then demonstrate that, for $0\leq p \leq d-1$,
the effective actions
$\Gamma^{(m)}_p$ and $\Gamma^{(m)}_{d-p-1}$  differ by a topological invariant. 
Section 3 is devoted to alternative proofs of some of the results of Section 2 
specifically for the $d=4$ case. 
Effective actions for massive super $p$-forms in four dimensions are studied in 
Section 4. In Section 5 we discuss the obtained results and sketch several generalisations. Four technical appendices are included.
Appendix A collects the properties of the Hodge-de Rham operator. Appendix B gives a summary of the results concerning massless $p$-forms in $d$ dimensions. The effective action of a massless three-form in $d=4$ is discussed in Appendix C. Finally, Appendix D 
describes dual formulations in the presence of a topological mass term.
We make use of the Grimm-Wess-Zumino geometry \cite{GWZ} which underlies the Wess-Zumino formulation \cite{WZ} for old minimal supergravity (see \cite{WB} for a review) discovered independently in \cite{Siegel77,old1,old2}.
Our two-component spinor notation and conventions follow \cite{BK}.
The algebra of the supergravity covariant derivatives, which we use, is given in Section 5.5.3 of \cite{BK}.

In order to have a uniform notation for non-supersymmetric and supersymmetric theories, in this paper we make use of the vielbein formulation for gravity. 
The background gravitational field  is  described by a vielbein $e^a = \rd x^m e_m{}^a (x)$, such that 
  $e=\det (e_m{}^a ) \neq 0$, and the metric is a composite field defined by $g_{mn} = e_m{}^a e_n{}^b \eta_{ab} $, with $\eta_{ab}$ the Minkowski metric. 
  All $p$-form fields in $d$ dimensions carry Lorentz indices. We make use of the torsion-free covariant derivatives
\bea
\nabla_a = e_a +\o_a = e_a{}^m  \pa_m +\hf \o_a{}^{bc}  M_{bc} ~, \qquad 
[\nabla_a , \nabla_b ] = \hf R_{ab}{}^{cd} M_{cd} ~.
\eea
Here $M_{bc} = -M_{cb}$ denotes the Lorentz generators, 
$ e_a{}^m (x) $ the inverse vielbein, $e_a{}^m e_m{}^b = \d_a{}^b$.  
The Lorentz generators act on a $d$-vector $v^a$ as 
$M_{bc} v^a = \d^a{}_b v_c - \d^a{}_c v_b = 2\d^a{}_{[b} b_{c]}$. 


\section{Massive $p$-forms in $d$ dimensions} \label{Section2}

In this section we derive effective actions  $\G^{(m)}_p $  for massive $p$-form models
in curved space and demonstrate that  
$\G^{(m)}_p $ and  $\G^{(m)}_{d-p-1} $ differ by a topological invariant. 


\subsection{Classical dynamics}

Let $B_{a_1 \dots a_p} (x)= B_{ [a_1 \dots a_p ]} (x) \equiv B_{a(p)} (x)$ be a differential $p$-form in curved space $\cM^d$. 
The dynamics of a massive $p$-form is described by the action 
\bea
S_p^{(m)} [B ] &=& -\frac{1}{2(p+1)! } \int \rd^d x \, e\, 
F^{a_1 \dots a_{p+1} } (B) F_{a_1 \dots a_{p+1}} (B)\non \\ 
&&- \frac{m^2}{2 p! } 
 \int \rd^d x \, e\, 
B^{a_1 \dots a_{p} } B_{a_1 \dots a_{p}}~,
\label{3.1} 
\eea
where $F_{a_1 \dots a_{p+1} } (B) := (p+1) \nabla_{[a_1} B_{a_2 \dots a_{p+1}] } $ is the field 
strength, and $m$ the mass. It is assumed in this section that $m\neq 0$.
The Euler-Lagrange equation corresponding to  \eqref{3.1}  is 
\bea
\nabla^b F_{b a_1 \dots a_p} (B) - m^2 B_{a_1 \dots a_p}=0 ~.
\eea
It implies that 
\bea
\nabla^c B_{c a_1 \dots a_{p-1} } =0 ~, 
\eea
and therefore the equation of motion turns into 
\bea
(\Box_p - m^2 ) B_{a_1 \dots a_p} =0~,
\eea
where $\Box_p$ is the covariant d'Alembertian \eqref{A.1}.

The symmetric energy-momentum tensor corresponding to the model \eqref{3.1} is 
\bea
T^{ab}_{[p,m]} (B)&=& \frac{1}{p!} \Big\{ F^{a c_1 \dots c_p} (B) F^b{}_{c_1 \dots c_p} (B)
-\frac{1} {2(p+1)} \eta^{ab}  F^{c_1 \dots c_{p+1}} (B) F_{c_1 \dots c_{p+1}} (B) \Big\} \non \\
&&+ \frac{m^2}{(p-1)!} \Big\{ B^{a c_1 \dots c_{p-1}}  B^b{}_{c_1 \dots c_{p-1}} 
-\frac{1} {2p} \eta^{ab}  B^{c_1 \dots c_{p}} B_{c_1 \dots c_p} \Big\}~,
\label{EMtensor}
\eea
with $\eta_{ab} $ the Minkowski metric.
It is conserved,
\bea
\nabla_b T^{ab}_{[p,m]} =0~,
\eea
on the  mass shell.


\subsection{Duality equivalence} 

It is known that the massless models for a $p$-form and $(d-p-2)$-form are classically equivalent, see Appendix \ref{AppendixB}.
In the massive case, however, a  $p$-form is dual to a $(d-p-1)$-form, 
see, e.g., \cite{Townsend,CFG}.
Here we recall the proof of this result. To demonstrate that the massive theories with actions 
$S_p^{(m)} [B ] $ and $S_{d-p-1}^{(m)} [A ] $ are equivalent, we first rewrite \eqref{3.1}
in the form 
\begin{subequations}
\bea
S_p^{(m)} [B ] &=& 
\frac{1}{2(d-p-1)! } 
\int \rd^d x \, e\, 
L^{a(d-p-1) } (B) L_{a(d-p-1)} (B)
\non \\ 
&&
- \frac{m^2}{2 p! } 
 \int \rd^d x \, e\, 
B^{a(p) } B_{a(p)} ~, 
\eea
where 
\bea
L^{a_1 \dots a_{d-p-1}} (B)&:=&\frac{1} {(p+1)!} \ve^{a_1 \dots a_{d-p-1} b_1 \dots b_{p+1} }
F_{b_1 \dots b_{p+1} } (B) \equiv * F^{a_1 \dots a_{d-p-1}} (B)~,
\eea
\end{subequations}
and introduce the  first-order action
\bea
S[B_p, L_{q}, A_{q}]&=&
\int \rd^d x \, e\, 
\bigg\{\frac{1}{2q! } 
L^{a(q) }  L_{a(q)} 
- \frac{m^2}{2 p! } 
B^{a(p) } B_{a(p)} \non \\
&&+\frac{m}{q!} A_{a(q) } \Big( L^{a(q) }- 
* F^{a (q)} (B) \Big) \bigg\} ~,\qquad q = d-p-1~.
\label{F-OM}
\eea
Here the variables $L_q$ and $A_q$ are unconstrained $(d-p-1)$-forms. 
Varying with respect to $A_{q}$ returns the original action, eq. \eqref{3.1}.
On the other hand, varying with respect to $L_{q}$ and $B_p$ leads to the dual action 
$S_{d-p-1}^{(m)} [A ] $.

The equations of motion corresponding to \eqref{F-OM} are
\begin{subequations}
\bea
mB_{a_1 \dots a_p}&=& -(-1)^{d(p+1)} * F_{a_1 \dots a_p} (A) ~, \\
m A_{a_1 \dots a_q} &:=&-L_{a_1 \dots a_{q}} ~, \\
L_{a_1 \dots a_q} &=&  *F_{a_1 \dots a_{q}} (B)~.
\eea
\end{subequations}
Making use of these equations, one may show that the energy-momentum tensors
in the theories $S_p^{(m)} [B ] $ and $S_{d-p-1}^{(m)} [A] $  coincide, 
\bea
T^{ab}_{[p,m]} (B) = T^{ab}_{[d-p-1,m]} (A)~.
\eea


\subsection{Quantisation}

Associated with the massive $p$-form model \eqref{3.1} is the effective action $\G^{(m)}_p $ defined by 
\bea
\re^{\ri \,\G^{(m)}_p } =  \int [ \cD B_p ] \, \re^{\ri S_p^{(m)} [B]} ~.
\label{3.5}
\eea
To obtain a useful expression for $\G^{(m)}_p$, we introduce a Stueckelberg  reformulation
of the theory. It is obtained from \eqref{3.1} by replacing 
\bea
B_{a_1 \dots a_p} ~\to ~ B_{a_1 \dots a_p } + \frac{1}{m} F_{a_1 \dots a_p} (V) ~,
\eea
for some $(p-1)$-form $V_{a_1 \dots a_{p-1} } (x) $. 
The resulting action 
\bea
S_p^{(m)} [B ,V] &=& -\frac{1}{2(p+1)! } \int \rd^d x \, e\, 
F^{a (p+1)  } (B) F_{a (p+1)} (B)- \frac{m^2}{2 p! } 
 \int \rd^d x \, e\, 
B^{a (p) } B_{a (p)} \non \\
&& -\frac{1}{2p! } \int \rd^d x \, e\, 
F^{a (p)  } (V) F_{a (p)} (V)
-\frac{m}{p! } \int \rd^d x \, e\, 
B^{a (p)  }  F_{a (p)} (V)
\label{3.7}
\eea
is invariant under gauge transformations 
\bea
\d_\z B_{a(p)} = p \nabla_{[a_1} \z_{a_2 \dots a_p]} \equiv F_{a(p)} (\z) ~, \qquad 
\d_\z V_{a(p-1)} = - m \z_{a(p-1)}~. 
\label{GF2.14}
\eea
The gauge freedom allows us to choose the gauge condition 
$ V_{a(p-1)} =0$ and then we are back to the original model.

The compensating field $V_{a(p-1)}$ appears in the action \eqref{3.7} only via 
the field strength $F_{a (p)} (V)$ which is invariant under gauge transformations 
\bea
\d_\l V_{a(p-1)} = (p-1) \nabla_{[a_1} \l_{a_2 \dots a_{p-1}]}
\equiv F_{a(p-1)} (\l) ~.
\label{extra}
\eea
This gauge freedom is characterised by linearly dependent generators, which makes it tempting  to conclude that the gauge theory under consideration is reducible. 
Nevertheless, \eqref{3.7} is an irreducible gauge theory and  can be quantised \`a la Faddeev and Popov. The point is that \eqref{extra} is a special case of the transformation 
\eqref{GF2.14} with $\z_{a(p-1)} = - m^{-1} F_{a(p-1)} (\l) $.

To quantise the gauge theory with action \eqref{3.7}, we choose the gauge fixing
\bea
\c_{a(p-1)} = \nabla^b B_{b a(p-1)} + m V_{a(p-1)} -\r_{a(p-1)}~,
\eea
with $\r_{a(p-1)}$ an external field. The gauge variation of $\c_{a(p-1)}$ is 
\bea
\d_\z \c_{a(p-1)} = \nabla^b F_{ba(p-1)} (\z) - m^2 \z_{a(p-1) } \equiv ({\mathfrak O} \z)_{a(p-1)} ~.
\eea
Here $\mathfrak O$ is the kinetic operator in the massive $p$-form model \eqref{3.1}. 
Making use of \eqref{3.5}, we conclude that the Faddeev-Popov determinant $\D_{\rm FP}$
is 
\bea
\D_{\rm FP} =\det {\mathfrak O} = \exp \Big( -2\ri \,\G^{(m)}_{p-1} \Big) ~.
\eea
Now, in accordance with the Faddeev-Popov procedure, the effective action is 
\bea
\re^{\ri \,\G^{(m)}_p } =  \int [ \cD B_p ] [\cD V_{p-1} ]\, \det {\mathfrak O} \,
\d \Big[\nabla^b B_{b a(p-1)} + m V_{a(p-1)} -\r_{a(p-1)}\Big] \, 
\re^{\ri S_p^{(m)} [B,V]} ~.
\eea
Averaging the right-hand side over $\r_{a(p-1)}$ with weight 
\bea
\exp \Big\{ - \frac{\ri } {2(p-1)!}  \int \rd^d x \, e\, 
\r^{a (p-1) } \r_{a (p-1)} \Big\}
\eea
leads to
\begin{subequations}
\bea
\re^{\ri \,\G^{(m)}_p } &=&  \int [ \cD B_p ] [\cD V_{p-1} ]\, \det {\mathfrak O} \,
\re^{\ri S_{\rm quant} [B,V] }~, \\
S_{\rm quant} [B,V] &=&  \frac{1}{2 p! } 
 \int \rd^d x \, e\, 
B^{a (p) } (\Box_p - m^2) B_{a (p)}  
+ S_{p-1}^{(m)} [V]~.
\eea
\end{subequations} 

As a result, for the effective action we obtain
\bea
\exp \Big\{ \ri \,\G^{(m)}_p \Big\}  
= \exp\Big\{ - \ri \,\G^{(m)}_{p-1} \Big\} \, \Big[ \det (\Box_p - m^2) \Big]^{-\hf} ~.
\label{3.15}
\eea 
This is a recurrence relation. It leads to a simple expression for the effective action
\bea
\G^{(m)}_p = \frac{\ri}{2} \sum_{k=0}^{p} (-1)^k \ln \det (\Box_{p-k} - m^2) 
= \frac{\ri}{2}  (-1)^p \sum_{k=0}^{p} (-1)^k \ln \det (\Box_{k} - m^2) 
~.
\label{3.16}
\eea 
In the $d=4$ case, this result agrees with \cite{BKP}.

The representation \eqref{3.16} is formal since each term on the right-hand side contains UV divergences. 
This issue is addressed by introducing a regularisation for the effective action, 
$(\G^{(m)}_{p})_{\rm reg} $. We will use the following prescription:
\bea
(\G^{(m)}_{p})_{\rm reg}  =-  \frac{\ri}{2} (-1)^p
\int_0^{\infty} \frac{ \rd s }{s^{1-\o}} \re^{-\ri (m^2 -\ri \ve) s} 
\sum_{k=0}^{p} (-1)^k  \int \rd^d x \, e\,  {\rm tr} \,{\mathfrak U}_{k} (x,x|s) ~,
\label{EA2.25}
\eea
with $\o, \ve \to +0$.
Here the right-hand side involves the (heat) kernel of 
the evolution operator 
${\mathfrak U}_{k} (s) =\exp (\ri s \Box_k ) $ acting on the space of $k$-forms.  
The kernel  of ${\mathfrak U}_{k} (s) $ is defined by 
\bea
{\mathfrak U}_{a (k)} {}^{a'(k)} (x,x'|s) = 
\re^{\ri s \Box_k}\, \d_{a(k)}{}^{a'(k)} (x,x') ~,
\eea
where the delta-function is 
\bea
\d_{a(k)}{}^{a'(k)} (x,x') = k! \d_{[a_1}{}^{a'_1} \dots \d_{a_k]}{}^{a'_k} 
e^{-1} \d^d(x-x') =k! \d_{a_1}{}^{[a'_1} \dots \d_{a_k}{}^{a'_k]} 
e^{-1} \d^d(x-x') ~,~~
\eea
such that 
\bea
\frac{1}{k!} \int \rd^d x' \, e(x')\,  \d_{a(k)}{}^{a'(k)} (x,x')\, \o_{a'(k)} (x') 
= \o_{a(k)} (x) ~,
\eea
for any $k$-form $\o $. In accordance with the definition of the delta-function,
the trace over Lorentz indices in \eqref{EA2.25} is 
\bea
{\rm tr} \,{\mathfrak U}_{k} (x,x|s) = \frac{1}{k!} {\mathfrak U}_{a (k)} {}^{a(k)} (x,x|s) ~.
\eea


\subsection{Quantum equivalence}

In $d$ dimensions, the model for a massive $p$-form is classically equivalent to that for a massive $(d-p-1)$-form. Let us analyse whether this equivalence extends to the quantum theory.
Our analysis will be based on the fact that the spaces of $p$-forms and $(d-p)$-forms are isomorphic, and the corresponding Hodge d'Alembertians are related to each other as follows 
\bea
*(\Box_p \,\o)  = \Box_{d-p}\, (*\o) ~,
\label{3.17}
\eea
where $\o$ is an arbitrary $p$-form.

Making use of the relations \eqref{3.16} and \eqref{3.17}, one may show that 
\bea
\G^{(m)}_{p} - \G^{(m)}_{d-p-1} =  (-1)^p {\mathfrak X}^{(m)} ~, \qquad 
{\mathfrak X}^{(m)} := \frac{\ri}{2} \sum_{k=0}^d (-1)^k
\ln \det (\Box_k - m^2)~.
\label{TI1}
\eea
There are two distinct cases. If the dimension of space-time is odd, $d=2n+1$,
the functional ${\mathfrak X}^{(m)} $ can be seen to vanish identically, 
\begin{subequations}
\bea
d=2n+1 \quad \implies \quad {\mathfrak X}^{(m)} &=& 0~.
\eea
In the even-dimensional case, $d=2n$,  ${\mathfrak X}^{(m)}$ can be rewritten in the form:
\bea
{\mathfrak X}^{(m)} &=& \ri \sum_{k=0}^{d/2 -1}  (-1)^k
\ln \det (\Box_k - m^2)
+ \frac{\ri}{2} (-1)^{d/2}
\ln \det (\Box_{d/2} - m^2)~. \label{2.29b}
\eea
\end{subequations}
This functional is no longer identically zero.  However, it turns out  to be a topological invariant in the sense that 
\bea
\frac{\d}{\d e_m{}^a (x) } \mathfrak{X}^{(m)} =0~.
\label{TI2}
\eea

In order to prove \eqref{TI2},
let us  consider the regularised version of  
$ {\mathfrak X}^{(m)} $ 
\begin{subequations}
\bea \label{TI3} 
({\mathfrak X}^{(m)})_{\rm reg}  &=&
\hf \int_0^{\infty} \frac{ \rd s }{s^{1-\o}} \re^{-\ri (m^2 -\ri \ve)s} \U(s)~, 
\qquad  \o, \ve \to +0~,
\eea
where
we have introduced the functional
\bea
\U(s) &=&-\ri  \sum_{k=0}^{d} (-1)^k  \int \rd^d x \, e\,  {\rm tr} \,{\mathfrak U}_{k} (x,x|s) ~.
\label{TI3.b}
\eea
\end{subequations}
Giving the gravitational field a small disturbance, the functional $\U(s)$ varies as 
\bea
\d \U(s) &=& - s  \sum_{k=0}^{d} (-1)^k  \int \rd^d x \, e\,  {\rm tr} 
\Big\{ (\rd \d \rd^\dagger + \d \rd^\dagger \rd)
\,{\mathfrak U}_{k} (x,x'|s) \Big\} \Big|_{x=x'}~.
\eea 
This variation may be rearranged by making use of the 
Ward identities
\begin{subequations}
\bea
(k+1) \nabla_{[a_1} {\mathfrak U}_{a_2 \dots a_{k+1} ]} {}^{a'(k)} (x,x'|s) 
&=& - \nabla_{b'} {\mathfrak U}_{a (k+1)} {}^{b'a'(k)} (x,x'|s) ~,\\
-\nabla^b {\mathfrak U}_{b a_1 \dots a_{k-1}} {}^{a'(k)} (x,x'|s) 
&=& k \nabla^{[a'_1}  {\mathfrak U}_{a (k-1)} {}^{a'_2 \dots a'_k]} (x,x'|s)~,
\eea 
in conjunction with the relations
\bea
 \int \rd^d x \, e \Big\{ \nabla^{b} \nabla_{[b} {\mathfrak U}_{a (k) ]} {}^{a'(k)} (x,x'|s) 
+
\nabla^{[a'_{k+1}} \nabla_{[a_{k+1}} {\mathfrak U}_{a (k) ]} {}^{a'(k)]} (x,x'|s) 
\Big\} \Big|\Big|
&=&0
~,~~~~~
\\
\int \rd^d x \, e \Big\{ k 
\nabla_{[a_1} \nabla^{b} {\mathfrak U}_{|b| a_2 \dots a_{k-1} ]} {}^{a'(k)} (x,x'|s) 
+\nabla_{b'}  \nabla^{b} {\mathfrak U}_{ba (k-1) } {}^{b' a'(k-1)} (x,x'|s)  \Big\}\Big|\Big|
&=&0~,~~~~~~
\eea
\end{subequations} 
where the double vertical bar means setting $x=x'$ and $a = a'$. Then one obtains 
\bea
\d \U(s) =0~,
\eea
which is equivalent to \eqref{TI2}.

Similar arguments may be used to show that $\U(s)$ is actually $s$-independent, 
\bea
\frac{\rd}{\rd s} \U(s) &=& -   \sum_{k=0}^{d} (-1)^k  \int \rd^d x \, e\,  {\rm tr} 
\Big\{ (\rd  \rd^\dagger +  \rd^\dagger \rd)
\,{\mathfrak U}_{k} (x,x'|s) \Big\} \Big|_{x=x'}
=0~.
\eea
For small values of $s$, 
it is well known that  the diagonal heat kernel  has the  asymptotic expansion
 \bea
\tr \, {\mathfrak U}_k(x,x| s) =  \frac{\ri}{ (4\p\ri  s)^{d/2}} 
 \sum_{n=0}^{\infty} (\ri s)^n \tr \, {\mathfrak a}_{k}^{(n)} (x,x)  ~,
 \eea
with $a_n(x,x) $ the Seeley-DeWitt coefficients. As a result, the topological invariant 
\eqref{TI3.b} takes the form 
\bea
\U = \frac{1}{ (4\p  )^{d/2}}  \sum_{k=0}^{d} (-1)^k  \int \rd^d x \, e\,
    \tr \, {\mathfrak a}_k^{(d/2)} (x,x)  ~,
 \eea
which is the heat kernel expression for the Euler characteristic, see, e.g., \cite{Rosenberg}. 

The above analysis is a variant of the famous heat kernel proofs of the Chern-Gauss-Bonnet theorem, see \cite{Rosenberg} for a review.


\section{Massive $p$-forms in four dimensions} \label{Section3}

In this section we will present alternative proofs of some results from the previous section in the $d=4$ case. The topological mismatch ${\mathfrak X}^{(m)}$ in \eqref{TI1}
will be ignored. 

\subsection{Two-form field}

The model for a massive two-form in curved space is described by the action 
\bea
{S}_2^{(m)} [B] = \hf \int \rd^4 x \, e\, 
\Big\{ L^a(B)  L_a(B) -\frac{m^2}{2}  B^{ab} B_{ab} \Big\} ~,
\label{two-form1}
\eea
where we have denoted
\bea
L^a(B) = \hf \ve^{abcd} \nabla_b B_{cd} = \frac{1}{6} \ve^{abcd} F_{bcd} (B)~.
 \label{two-form2} 
\eea
This theory is classically equivalent to the model with action $S_1^{(m)} [V]$,  which describes the massive vector field in curved space. 

We are going to show that 
\bea
\exp \Big( \ri \G_2^{(m)} \Big) = \exp \Big( \ri \G_1^{(m)} \Big)~.
\label{two-form3}
\eea
For this we consider the following change of variables\footnote{Given an arbitrary  
$p$-form $\o_p$
on a compact Riemannian manifold $(M, g)$, the Hodge decomposition theorem states that 
$\o_p= \rd \vf_{p-1} + \rd^\dagger \J_{p+1} + h_p$, where $h_p$ is harmonic, 
$\Box_p h_p =0$. It is assumed in \eqref{two-form4.a} that  $\Box_p$ has no normalised zero modes. 
} 
\begin{subequations} \label{two-form4} 
\bea
B_{ab}&=& 2 \nabla_{[ a} V_{b]} + \ve_{abcd} \nabla^c K^d ~,  \label{two-form4.a} \\
\F &=& \nabla_a K^a ~, \qquad \J = \nabla_a V^a ~.
\eea
\end{subequations}
Its Jacobian proves to be
\bea
J(B, \F , \J| V, K ) = \det \Box_1 ~.
 \label{two-form5} 
\eea
We perform the change of variables \eqref{two-form4} in the action 
\bea
S[B, \F, \J]  = S_2^{(m)} [B]  -\hf \int \rd^4 x \, e\, \F (\Box_0 -m^2) \F
- \frac{m^2}{2} \int \rd^4 x \, e\, \J^2 ~.
\eea
Then $S[B, \F, \J]$ turns into
\bea
\frac{m^2}{2} \int \rd^4 x \, e\, V^a \Box_1 V_a 
+ \hf \int \rd^4 x \, e\, K^a ( \Box_1 - m^2 ) \Box_1 K_a~.
\eea 
Making use of \eqref{two-form5} leads to 
\bea
\exp \Big( \ri \G_2^{(m)} \Big) \Big[\det (\Box_0 -m^2) \Big]^{-1/2}  
&=& \int [\cD B_2] [\cD \F ] [\cD \J] \exp \Big( \ri S[B, \F, \J]  \Big) \non \\
&=& \Big[\det (\Box_1 -m^2)\Big]^{-1/2} ~,
\eea
which is equivalent to \eqref{two-form3}.


\subsection{Three-form field}

The model for a massive three-form in curved space is described by the action 
\bea
{S}_3^{(m)} [V] = \hf \int \rd^4 x \, e\, 
\Big\{ (\nabla_a V^a)^2 +m^2 V^a V_a \Big\} ~.
\label{2.11}
\eea
In terms of the field strength $H =\nabla_a V^a$, the equation of motion is 
\bea
\nabla_a H -m^2 V_a =0 \quad \implies \quad (\Box_0 - m^2 ) H =0~.
\eea
This shows that the three-form model \eqref{2.11} is equivalent to 
the massive scalar model
\bea
{S}_0^{(m)} [\vf] =  - \hf \int \rd^4 x \, e\, 
\Big\{ \nabla^a \vf \nabla_a \vf +m^2 \vf^2 \Big\} ~.
\label{2.13}
\eea
Classical equivalence of the theories \eqref{2.11} and \eqref{2.13} 
is established by considering a first-order model with Lagrangian
\bea
\cL = \r \nabla_a V^a -\hf \r^2 + \hf m^2 V^a V_a ~.
\eea

The effective action for the massive three-form model is
\bea
\exp \Big( \ri \G_3^{(m)} \Big) = \int [ \cD V ] \, \re^{\ri S_3^{(m)} [V]} ~.
\eea 
We are going to show that 
\bea
\exp \Big( \ri \G_3^{(m)} \Big) = \exp \Big( \ri \G_0^{(m)} \Big)~.
\label{2.16}
\eea
For this we consider the following change of variables \cite{BK88} 
\begin{subequations} \label{2.17}
\bea
V_a &=& \nabla_a \F + \hf \ve_{abcd} \nabla^b B^{cd} 
\equiv \nabla_a \F + L_a (B) ~, \\
K_a &=& \nabla_a \J + \nabla^b B_{ab} 
\equiv \nabla_a \J + G_a (B) ~.
\eea
\end{subequations}
The corresponding Jacobian  is
\bea
J( V,K| B, \F, \J) = \det \Box_0 \, (\det \Box_2 )^\hf~,
\label{2.18}
\eea
see \cite{BK88} for the derivation.
We perform the above change of variables in the path integral 
\bea
\exp \Big( 2 \ri \G_3^{(m)} \Big) = \int [ \cD V ]  [ \cD K ]\, 
\exp  \ri \Big(  S_3^{(m)} [V] - S_3^{(m)} [K] \Big)
\eea 
For the action 
${S}_3^{(m)} [V] - {S}_3^{(m)} [K] $ we obtain
\bea
{S}_3^{(m)} [V] - {S}_3^{(m)} [K] 
&=& \frac 14 m^2  \int \rd^4 x \, e\, B^{ab}\Box_2 B_{ab}
+ \hf  \int \rd^4 x \, e\, \F \Box_0 ( \Box_0 - m^2 ) \F \non \\
&&-\hf  \int \rd^4 x \, e\, \J \Box_0 ( \Box_0 - m^2 ) \J~.
\eea
Then, taking into account \eqref{2.18} leads to \eqref{2.16}.


\section{Massive super $p$-forms in four dimensions} \label{Section4}

In this section we  study effective actions of the following massive locally supersymmetric theories in four dimensions: (i) vector multiplet; (ii)  tensor multiplet; 
and (iii) three-form multiplet. In the massless case, these multiplets are naturally described 
in terms of super $p$-forms, with $p =1, 2$ and 3, respectively. The models for massive vector and tensor multiplets are classically equivalent. Here we will demonstrate their quantum equivalence. 

At the component level, the locally supersymmetric models of our interest contain
the massive $p$-form models we have studied in the previous section. 

\subsection{Setup} 

The massive vector multiplet in a supergravity background 
\cite{Siegel-tensor,VanProeyen} is described in terms of a real scalar prepotential $V$.  
The action is 
\bea
S^{(m)}_{\rm vector} [V]= \hf \int\rd^4x\rd^2\q\rd^2\qb\, E \, V \Big\{ \frac{1}{8}
 \cD^\a (\bar \cD^2-4R) \cD_\a + m^2 \Big\} V~, \qquad \bar V = V~.
 \label{4.1}
 \eea
The massive tensor multiplet \cite{Siegel-tensor}
is described in terms of a covariantly chiral spinor superfield 
$\J_\a$, $\bar \cD_\bd \J_\a =0$, and its conjugate $\bar \J_\ad$. The action is 
\bea
S^{(m)}_{\rm tensor} [\J, \bar \J] = -\hf \int\rd^4x\rd^2\q\rd^2\qb\, E \, \big( G(\J)\big)^2 - \bigg\{ \frac{m^2}{2}  \int\rd^4x\rd^2\q\, \cE \,\J^2
+{\rm c.c.} \bigg\}~,
\label{4.2}
\eea
where we have introduced the real superfield
\bea
G(\J) := \hf \big( \cD^\a \J_\a + \bar \cD_\ad \bar \J^\ad \big) ~,
\label{G}
\eea
which is covariantly linear, $(\bar \cD^2 - 4R) G =0$. 
Similar to the vector multiplet, the massive three-form multiplet is formulated 
in terms of a real scalar prepotential $V$. The corresponding action is obtained from 
\eqref{1.3} by adding a mass term, 
\bea
S^{(m)}_{\text{3-form}} [V]= \hf \int\rd^4x\rd^2\q\rd^2\qb\, E \, V\Big\{ \big(\cP_+ + \cP_- \big)^2 
- m^2 \Big\}V ~,
\label{4.3}
\eea
where the operators $\cP_+ $ and $ \cP_- $ are defined in \eqref{4.4}.
We recall that  $\cP_+ U$ and  $\cP_- U$ are covariantly chiral and
 antichiral, respectively, for any scalar superfield $U$.

Associated with the above massive models are their effective actions 
defined by 
\begin{subequations}\label{4.6}
\bea
\re^{ \ri \G^{(m)}_{\rm vector}} &=& \int [\cD V] \,\re^{ \ri S^{(m)}_{\rm vector} [V] }~, 
\label{4.6a} \\
\re^{ \ri \G^{(m)}_{\rm tensor}} &=& \int [\cD \J] [\cD \bar \J] \,
\re^{ \ri S^{(m)}_{\rm tensor} [\J, \bar \J] }~,  \label{4.6b} \\
\re^{ \ri \G^{(m)}_{\text{3-form}} }&=& \int [\cD V] \,\re^{ \ri S^{(m)}_{\text{3-form}} [V] }~.
\label{4.6c}
\eea
\end{subequations}
There exist  alternative representations for the effective actions introduced. They may be derived by making use of Stueckelberg reformulations of the models under consideration.


\subsection{Quantisation of the massive vector multiplet model}

The Stueckelberg reformulation of the massive vector multiplet model is obtained by replacing 
\bea
V ~\to ~ V + \frac{1}{m} \big( \cP_+ + \cP_- \big) K ~, \qquad \bar K = K
\label{4.7} 
\eea
in the action \eqref{4.1}. The resulting action 
\bea
S^{(m)}_{\rm vector} [V,K]&=& \hf \int\rd^4x\rd^2\q\rd^2\qb\, E \, \bigg\{ \frac{1}{8}
 V\cD^\a (\bar \cD^2-4R) \cD_\a V + m^2  V^2 \non \\
&&  +2 m V \big( \cP_+ + \cP_- \big) K + K \big( \cP_+ + \cP_- \big)^2 K \bigg\}
\label{4.8}
 \eea
is invariant under gauge transformations
\bea
\d_U V = \big( \cP_+ + \cP_- \big) U ~, \qquad \d_U K = - m U~,  \qquad \bar U = U~.
\eea

To quantise the gauge theory with action \eqref{4.8}, we introduce the gauge fixing 
\bea
\c = \big( \cP_+ + \cP_- \big) V + m K - {\mathfrak F}~, 
\eea
with ${\mathfrak F}$ a background real superfield. The gauge variation of $\c$ is 
\bea
\d_U \c =  \big( \cP_+ + \cP_- \big)^2 U  - m^2 U \equiv {\mathfrak O}U~,
\eea 
and therefore the Faddeev-Popov determinant is 
\bea
\D_{\rm FP} = \det {\mathfrak O} = \exp \Big( - 2\ri \G^{(m)}_{\text{3-form}} \Big)~.
\eea
For the effective action we obtain 
\bea
\re^{ \ri \G^{(m)}_{\rm vector}} &=& \int [\cD V] [\cD K ]\,\det {\mathfrak O} \,
\d\Big[\big( \cP_+ + \cP_- \big) V + m K - {\mathfrak F}\Big]
\re^{ \ri S^{(m)}_{\rm vector} [V,K] }~.
\eea
Averaging the right-hand side over ${\mathfrak F}$ with weight
\bea
\exp \Big\{ -\frac{\ri } {2} \int\rd^4x\rd^2\q\rd^2\qb\, E \, {\mathfrak F}^2\Big\} ~,
\eea
we obtain 
\bea
\exp \Big\{  \ri \G^{(m)}_{\rm vector} \Big\}  = \exp \Big\{ - \ri \G^{(m)}_{\text{3-form}} \Big\} 
\Big[ \det \big(\Box^{(R)}_{\rm v} - m^2 \big)\Big]^{-\hf} ~,
\label{4.15}
\eea
where we have introduced the operator\footnote{The d'Alembertian  $\Box^{(R)}_{\rm v}$
is a member of the family of operators $\Box^{(\F)}_{\rm v}$ introduced in \cite{BK86,BK},
where 
$\Box^{(\F)}_{\rm v} =-\frac 18  \cD^\a (\bar \cD^2-4R) \cD_\a 
+ \big\{ \cP_+ , \cP_- \big\} + \F \cP_+ + \bar \F \cP_-$, with $\F $ a chiral scalar.
}
\cite{BK86,BK}
\bea
\Box^{(R)}_{\rm v} &=& -\frac 18  \cD^\a (\bar \cD^2-4R) \cD_\a 
+ \big( \cP_+ + \cP_- \big)^2
\non \\
&=& \cD^a \cD_a -\frac 14 G^{\a\ad} \big[ \cD_\a , \bar \cD_\ad \big]
- \frac{1}{4} R (\bar \cD^2 -4R) - \frac{1}{4} \bar R ( \cD^2 - 4 \bar R) \non \\
&&-\frac 14 (\cD^\a R) \cD_\a - \frac 14 (\bar \cD_\ad \bar R) \bar \cD^\ad 
-\frac 14 (\cD^2R) - \frac 14 (\bar \cD^2 \bar R) + 2 R \bar R~. 
\eea
Our final result \eqref{4.15} relates the effective actions \eqref{4.6a} and \eqref{4.6c}.


\subsection{Quantisation of the massive tensor multiplet model}

The Stueckelberg reformulation of the massive tensor multiplet model, 
eq. \eqref{4.2},  is obtained by replacing 
\bea
\J_\a ~\to ~ \J_\a + \frac{\ri}{2m} W_\a(V) ~, \qquad 
W_\a = -\frac 14 (\bar \cD^2 -4R) \cD_\a V~, \qquad \bar V = V
\eea
in the action \eqref{4.2}. This leads to the action
\bea
S^{(m)}_{\rm tensor} [\J, \bar \J, V] &=& -\hf \int\rd^4x\rd^2\q\rd^2\qb\, E \, \big( G(\J)\big)^2
 - \bigg\{ \frac{m^2}{2}  \int\rd^4x\rd^2\q\, \cE \,\J^2
+{\rm c.c.} \bigg\} \non \\
&& +  \int\rd^4x\rd^2\q\rd^2\qb\, E \, \Big\{\frac{1}{16} V 
 \cD^\a (\bar \cD^2-4R) \cD_\a V +m 
  V L(\J) \Big\}~,
\label{4.18}
\eea
where we have introduced the covariantly linear superfield
\bea
L(\J) := \frac{\ri}{2} \big( \cD^\a \J_\a - \bar \cD_\ad \bar \J^\ad \big)~.
\eea
The action is invariant under gauge transformations 
\bea
\d_K \J_\a = -\frac{\ri}{8} \big( \bar \cD^2 - 4R \big) \cD_\a K ~, \qquad 
\d_K V = - m K~, \qquad \bar K =K~.
\eea

To quantise the gauge theory with action \eqref{4.18}, we introduce the gauge fixing 
\bea
\c = L(\J) - m V - {\mathfrak U}~,
\eea
where ${\mathfrak U}$ is a background real superfield. The gauge variation of $\c$ is 
\bea
\d_K \c = \frac{1}{8}\cD^\a \big( \bar \cD^2 - 4R \big) \cD_\a K +m^2 K \equiv {\mathfrak O} K ~.
\eea
Here ${\mathfrak O} $ is exactly the operator which determines the vector multiplet action 
\eqref{4.1}. This means that the Faddeev-Popov determinant is 
\bea
\D_{\rm FP} = \det {\mathfrak O} = \exp \Big( - 2\ri \G^{(m)}_{\text{vector}} \Big)~.
\eea
As a result, the effective action can be written in the form 
\bea
\re^{ \ri \G^{(m)}_{\rm tensor}} &=& \int [\cD \J ] [ \cD \bar \J] [\cD V] \,\det {\mathfrak O} \,
\d\Big[ L(\J) - m V - {\mathfrak U} \Big]
\re^{ \ri S^{(m)}_{\rm tensor} [\J, \bar \J, V] }~.
\label{4.24}
\eea

Since the right-hand side of \eqref{4.24} is independent of ${\mathfrak U}$, 
we can average it over ${\mathfrak U}$ with weight 
\bea
\exp \Big\{ \frac{\ri } {2} \int\rd^4x\rd^2\q\rd^2\qb\, E \, {\mathfrak U}^2\Big\} ~.
\eea
This leads to 
\bea
\exp \Big\{  \ri \G^{(m)}_{\rm tensor} \Big\}  = \exp \Big\{ - \ri \G^{(m)}_{\text{vector}} \Big\} 
\Big[ \det \big({\Box}_{\rm c} - m^2 \big)  \det \big({\Box}_{\rm a} - m^2 \big) \Big]^{\hf} ~,
\label{4.26}
\eea
where the d'Alembertian $\Box_{\rm c}$  acts on the space of covariantly chiral spinors
\cite{GNSZ,BK88}
\bea
\Box_{\rm c} \J_\a &:=& \frac{1}{16} \big( \bar \cD^2 -4R\big) \big( \cD^2 - 6 \bar R \big) \J_\a 
\non \\
&=&\Big\{ \cD^b \cD_b + \frac 14 R \cD^2 + \ri G^b \cD_b + \frac 14 (\cD^\b R) \cD_\b
- \frac 38 \big( \bar \cD^2 -4R\big) \bar R \Big\}\J_\a \non \\
&& - \Big\{ W^\b{}_{\a \g } \cD_\b +\hf (\cD^\b W_{\a\b\g}) \Big\} \J^\g~.
\eea
Our final result \eqref{4.26} relates the effective actions \eqref{4.6a} and \eqref{4.6b}.


\subsection{Quantisation of the massive three-form multiplet model}

The Stueckelberg reformulation of the massive tensor multiplet model, 
eq. \eqref{4.3},  is obtained by replacing 
\bea
V ~ \to ~ V +\frac{1}{m} G(\J) ~, \qquad G(\J) := \hf \big( \cD^\a \J_\a + \bar \cD_\ad \bar \J^\ad \big) ~, \qquad \bar \cD_\bd \J_\a=0~.
\eea
The resulting action 
\bea
S^{(m)}_{\text{3-form}} [V, \J, \bar \J]
&=& \hf \int\rd^4x\rd^2\q\rd^2\qb\, E \, V\Big\{ \big(\cP_+ + \cP_- \big)^2 
- m^2 \Big\}V \non \\
&& -   \int\rd^4x\rd^2\q\rd^2\qb\, E \, \Big\{ m V G(\J) + \hf \big( G(\J) \big)^2 \Big\}
\label{4.29}
\eea
is invariant under gauge transformations
\bea
\d_\l V = \hf \big( \cD^\a \l_\a + \bar \cD_\ad \bar \l^\ad \big) ~, \qquad 
\d_\l \J_\a = - m \l_\a ~, \qquad \bar \cD_\bd \l_\a =0~.
\eea

To quantise the gauge theory with action \eqref{4.29}, we introduce the gauge  condition 
\bea
\c_\a = 
\hf W_\a(V) 
+ m \J_\a - \x_\a ~, \qquad 
W_\a(V) := -\frac 14 \big(\bar \cD^2 - 4R \big) \cD_\a V ~,
\label{4.31}
\eea
where $\x_\a $ is a background chiral spinor.  The gauge variation of $\c_\a$ is 
\bea
\d_\l \c_\a = - \frac 18 \big(\bar \cD^2 - 4R \big) \cD_\a G(\l) - m^2 \l_\a 
\equiv {\mathfrak O} \l_\a~.
\eea
Here $\mathfrak O$ is the operator which determines the massive tensor multiplet model 
\eqref{4.2}. This means that the Faddeev-Popov {\it super-determinant} is 
\bea
\D_{\rm FP} = [\det {\mathfrak O}]^{-1} =  \exp \Big( - 2\ri \G^{(m)}_{\text{tensor}} \Big)~.
\eea
Therefore, the effective action is given by the path integral
\bea
\re^{ \ri \G^{(m)}_{\text{3-form}}} &=& \int [\cD V] [\cD \J ] [ \cD \bar \J]  \,\det {\mathfrak O} \,
\re^{ \ri S^{(m)}_{\text{3-form}} [V,\J, \bar \J] } \non \\
&& \times \d\Big[  \frac 12 W_\a(V) + m \J_\a - \x_\a  \Big]
\d\Big[ \hf \bar W_\ad(V) + m \bar \J_\ad - \bar \x_\ad  \Big]
\label{4.34}
\eea
Since the right-hand side is independent of the chiral spinor $\x_\a$ and its conjugate
$\bar \x_\ad$, we can average over these superfields with weight 
\bea
\exp \Big\{- \frac{\ri}{2} \Big(  \int\rd^4x\rd^2\q\, \cE \,\x^2
+{\rm c.c.} \Big) \Big\}~.
\eea
This will lead to the relation
\bea
\exp \Big\{  \ri \G^{(m)}_{\text{3-form}} \Big\}  = \exp \Big\{ - \ri \G^{(m)}_{\text{tensor}} \Big\} 
\Big[ \det \big(\Box^{(R)}_{\rm v} - m^2 \big)\Big]^{-\hf} ~,
\label{4.36}
\eea
which connects the effective actions \eqref{4.6b} and \eqref{4.6c}.


\subsection{Analysis of the results} 

We have derived three different relations which connect the three effective actions 
defined in \eqref{4.6}. They are given by the equations \eqref{4.15}, \eqref{4.26} and 
\eqref{4.36}. These results have nontrivial implications. 
Firstly, it follows from \eqref{4.15} and \eqref{4.36} that 
\bea
\G^{(m)}_{\text{vector}} = \G^{(m)}_{\text{tensor}} ~.
\label{4.37} 
\eea
Therefore, the classically equivalent theories \eqref{4.1} and \eqref{4.2} 
remain equivalent at the quantum level. Secondly, making use of \eqref{4.26} and \eqref{4.37}
leads to 
\bea
\G^{(m)}_{\text{vector}} = \G^{(m)}_{\text{tensor}}  
= -\frac{\ri}{4} \ln \det \big({\Box}_{\rm c} - m^2 \big) 
-\frac{\ri}{4}  \ln \det \big({\Box}_{\rm a} - m^2 \big) ~.
\label{4.38}
\eea
Thirdly, from \eqref{4.36} and \eqref{4.38} we deduce 
\bea
 \G^{(m)}_{\text{3-form}}  = \frac{\ri}{2} \ln \det \big(\Box^{(R)}_{\rm v} - m^2 \big)
+ \frac{\ri}{4} \ln \det \big({\Box}_{\rm c} - m^2 \big) 
+\frac{\ri}{4}  \ln \det \big({\Box}_{\rm a} - m^2 \big) ~.
\label{4.39}
\eea
The superfield heat kernels corresponding to the operators appearing in \eqref{4.38} and \eqref{4.39} 
were studied in \cite{BK86,BK88,BK,McA}.

As follows from  \eqref{4.38},  the effective actions $\G^{(m)}_{\text{vector}}$ and $ \G^{(m)}_{\text{tensor}}  $ coincide, without any topological mismatch.
This is due to the  use of the Stueckelberg formulation defined by eqs. \eqref{4.7} 
and \eqref{4.8}. A topological mismatch will emerge if we  consider a slightly 
different Stueckelberg reformulation, which is obtained by replacing the dynamical superfield in  \eqref{4.1} by the rule
\bea
V ~\to ~ V + \frac{1}{m} \big( \F + \bar \F \big)~, \qquad \bar \cD_\ad \F =0.
\eea
 This leads to the action 
\bea
S^{(m)}_{\rm vector} [V,\F, \bar \F]&=& \hf \int\rd^4x\rd^2\q\rd^2\qb\, E \, \bigg\{ \frac{1}{8}
 V\cD^\a (\bar \cD^2-4R) \cD_\a V + \big( m  V + \F +\bar \F \big)^2  \bigg\}~,~~
  \eea
  which possesses the gauge freedom 
\bea
  \d_\l V = \l +\bar \l ~, \qquad \d_\l \F = -m \l~, \qquad \bar \cD_\ad \l =0~.
\eea

Modulo a purely topological contribution, the functional \eqref{4.39} proves to be twice 
the effective action of a scalar multiplet. To justify this claim, let us consider the following dynamical system
\bea
S^{(m)} [V, \F, \bar \F]
&=& \hf \int\rd^4x\rd^2\q\rd^2\qb\, E \, \bigg\{ V\big(\cP_+ + \cP_- \big)^2 V
+\big(\F + \bar \F \big)^ 2 + 2m V \big( \F + \bar \F\big) \bigg\} ~,~~~
\label{5.66}
\eea
where $\F$ is a chiral scalar. This model proves to be dual to the massive three-form 
theory \eqref{4.3}. The action \eqref{5.66} is invariant under gauge transformations 
\bea
\d_\l V = \hf \big( \cD^\a \l_\a + \bar \cD_\ad \bar \l^\ad \big) ~, 
\qquad \bar \cD_\bd \l_\a =0
\eea
corresponding to the massless three-form multiplet. Quantisation of the reducible gauge theory can be carried out using the method described in \cite{BK88}.
Next, we represent the chiral scalar $\F$ in \eqref{5.66} as
\bea
\F = \cP_+ U~, \qquad \bar U =U~.
\eea
Finally, we introduce new variables $K_\pm = \frac{1}{\sqrt{2} }( V\pm U)$. Then the action turns into
\bea
S^{(m)} [K_\pm ]
&=& \hf \int\rd^4x\rd^2\q\rd^2\qb\, E \, \bigg\{
 K_+ \big(\cP_+ + \cP_- \big)^2 K_+ +  K_- \big(\cP_+ + \cP_- \big)^2 K_-
\non \\
&&+m K_+\big(\cP_+ + \cP_- \big)K_+ -m K_-\big(\cP_+ + \cP_- \big)K_-\bigg\}~.
\eea
This is the three-form counterpart of the theory
\bea
S^{(m)}_{\rm scalar} [\F_\pm, \bar \F_\pm ]
&=& \hf \int\rd^4x\rd^2\q\rd^2\qb\, E \, \bigg\{  \big(\F_+ + \bar \F_+ \big)^2 
+\big(\F_- + \bar \F_- \big)^2\bigg\}
\non \\
&&+ \frac{m}{2} \bigg\{   \int\rd^4x\rd^2\q\, \cE \,\Big( \F_+^2 - \F_-^2\Big) 
+{\rm c.c.}   \bigg\}~,
\eea
which describes two decoupled massive scalar multiplets in a supergravity background.
The quantum effective action for this theory is
\bea
\G_{\rm scalar}^{(m)} = \frac{\ri}{2} \ln \Big({\rm Det} \,H^{(R+m)} {\rm Det} \,H^{(R-m)} \Big)~,
\label{4.47}
\eea
where $H^{(\j)} $ denotes the following operator \cite{BK86,BK}
\bea
H^{(\j)} =  \left(
\begin{array}{cc}
 \j   &  \cP_+ \ \\
 \cP_-  &   \bar \j
\end{array}
\right)~,
 \qquad \bar \cD_\ad \j =0 ~. 
 \eea
By definition, the operator $H^{(\j)}$  acts on the space  of chiral-antichiral column-vectors 
\bea
H^{(\j)} \left(
\begin{array}{c}
\eta  \\
 \bar \eta
\end{array}
\right) 
= \left(
\begin{array}{c}
\j \eta + \cP_+\bar \eta  \\
\bar \j \bar \eta + \cP_- \eta 
\end{array}
\right) ~, \qquad  \bar \cD^\ad \eta =0 ~.
\eea
A useful expression for ${\rm Det} \,H^{(\j)} $ in terms of the functional determinants 
of covariant d'Alembertians is derived in \cite{BK86,BK}.

Since the effective actions \eqref{4.39} and \eqref{4.47} should differ only by a topological term, we conclude that 
\bea
-2X^{(m)} &=&  \ri \ln \det \big(\Box^{(R)}_{\rm v} - m^2 \big)
+ \frac{\ri}{2} \ln \det \big({\Box}_{\rm c} - m^2 \big) 
+\frac{\ri}{2}  \ln \det \big({\Box}_{\rm a} - m^2 \big) \non \\
&&-\ri \ln \Big({\rm Det} \,H^{(R+m)} {\rm Det} \,H^{(R-m)} \Big)
\eea
is a topological invariant. It is a generalisation of the invariant introduced in 
\cite{GNSZ,BK88}.

Our analysis in this section provides the supersymmetric completion of the results obtained in Section \ref{Section3}.


\section{Discussion and generalisations} \label{Section5}

In this paper we derived compact expressions for the massive $p$-form effective actions 
for $0\leq p \leq d-1$, where $d$ is the dimension of curved spacetime.  
We then demonstrated  that the effective actions
$\Gamma^{(m)}_p$ and $\Gamma^{(m)}_{d-p-1}$  differ by a topological invariant. 
These results were extended to the case of massive super $p$-forms coupled to background ${\cal N}=1$ supergravity in four dimensions.
There are several interesting $p$-form models which we have not considered in this work and which deserve further studies. Here we briefly discuss such models. 

As a natural generalisation of the Cremmer-Scherk model for massive spin-1 in $d=4$
 \cite{CS}, the dynamics of a massive 
$p$-form in $d$ dimensions can be described in terms of a gauge-invariant action
involving two fields $B_p$ and $A_q$, with $q = d-p-1$, and  
a topological ($B\wedge F$) mass term. The action is  
\bea
S^{(m)} [B_p, A_{q}  ] &=& -\hf  \int \rd^d x \, e\, \bigg\{
\frac{1}{(p+1)! }F^{a (p+1) } (B) F_{a (p+1)} (B)\non \\
&& +\frac{1}{(q+1)! } 
F^{a(q+1) } (A) F_{a (q+1)} (A)
\bigg\}
+I^{(m)}[B_p, A_{q}  ] ~,
\label{5.1}
\eea
where $I^{(m)}$ stands for the topological mass term 
\bea
I^{(m)} [B_p, A_{q}  ] &=& \frac{m}{q! (p+1)!} \int \rd^d x \, e\,
\ve^{ a(q) b(p+1)} A_{a(q)}F_{b(p+1) } (B) \non \\
&=& (-1)^{d(d-p)}\frac{  m}{p! (d-p)! } \int \rd^d x \, e\,
\ve^{ a(p) b(q+1)} B_{a(p)}F_{b(q+1) } (A)~.
\eea
As is well known, 
this model is  dual to the massive theories  $S^{(m)}_p [B]$ and $S^{(m)}_q [A]$,  
with $S^{(m)}_p [B]$ defined by eq. \eqref{3.1}. The corresponding duality transformations 
are described, for completeness, in Appendix \ref{AppendixD}.

The action \eqref{5.1}  is invariant under gauge transformations
\bea
\d_\z B_{a(p)} = p \nabla_{[a_1} \z_{a_2 \dots a_p]} ~, \qquad 
\d_\x A_{a(q)} = q \nabla_{[a_1} \x_{a_2 \dots a_{q}]} ~.
\eea
The corresponding generators are linearly dependent,  and therefore
the  gauge theory \eqref{5.1} should be quantised using the Batalin-Vilkovisky formalism \cite{BV} or the simpler quantisation schemes \cite{Siegel,Obukhov,BK88},
which are specifically designed to quantise Abelian gauge theories. 
It would be interesting to show
that the effective action for the  gauge theory \eqref{5.1} coincides with \eqref{3.16}
modulo a topological invariant. 

In four dimensions, a supersymmetric generalisation of  the Cremmer-Scherk model 
was given by Siegel \cite{Siegel-tensor}
\bea
S^{(m)} [\J, \bar \J, V] &=& -\hf \int\rd^4x\rd^2\q\rd^2\qb\, E \, \Big\{ \big( G(\J)\big)^2 
-\frac{1}{8} V 
 \cD^\a (\bar \cD^2-4R) \cD_\a V \Big\} +I^{(m)}~,~~
 \label{5.33}
\eea
where the mass term is given by 
\bea
I^{(m)} [\J, \bar \J, V] &=& m  \int\rd^4x\rd^2\q\rd^2\qb\, E \, V  G(\J)
= -  \hf  m  \int\rd^4x\rd^2\q\, \cE \,\J^\a W_\a(V) 
+{\rm c.c.}
\label{5.55}
\eea
This is a dual formulation for the models \eqref{4.1} and \eqref{4.2}.
The action \eqref{5.33} is invariant under combined gauge transformations corresponding  
to the massless vector and tensor multiplets. This reducible massive gauge theory can be 
quantised using the method described in \cite{BK88}. 

The mass term \eqref{5.55} is locally superconformal \cite{CFG}. For the supergravity formulation used in the present paper,  this means that \eqref{5.55} is 
 super-Weyl invariant. We recall that a super-Weyl transformation of the covariant derivatives \cite{Siegel78,HT} is
\bea
\d_\S \cD_\a &=& ( {\bar \S} - \hf \S)  \cD_\a + \cD^\b \S \, M_{\a \b}  ~, \qquad
\d_\S \bar \cD_\ad  =  (  \S -  \hf {\bar \S})
\bar \cD_\ad +  ( \bar \cD^\bd  {\bar \S} )  {\bar M}_{\ad \bd} ~,
\eea
where the parameter $\S$ is chiral, $\bar \cD_\ad \S=0$, 
and $M_{\a\b}$ and $\bar M_{\ad \bd} $ are the Lorentz generators 
defined as in \cite{BK}. Such a transformation acts  on the prepotentials $V$ and $\J_\a$ as
\bea
\d_\S V =0~, \qquad \d_\S \J_\a = \frac 32 \S \J_\a~,
\eea
see \cite{BK} for the technical details. The mass term \eqref{5.55} is the supersymmetric version of the $d=4$ Green-Schwarz anomaly  cancellation term.

Another supersymmetric analogue of the Cremmer-Scherk model is described by the action
\eqref{5.66}.

If $d$ is even, $d=2n$, one can introduce massive $n$-form models with two types of mass terms \cite{QT,DQT,DF}, 
\bea
S^{(m,e)}_n [B  ] &=& -\frac{1}{2n!} \int \rd^d x \, e\, \bigg\{
\frac{1}{n+1 }F^{a (n+1) } (B) F_{a (n+1)} (B)\non \\
&& +m^2 B^{a(n)} B_{a(n)} + m e B^{a(n)} * B_{a(n)} \bigg\}~,
\label{5.4}
\eea
with $m$ and $e$ constant parameters. 
Here the second mass term vanishes if $n $ is odd 
(however, it is non-zero in the case of several $n$-forms \cite{DF}.)
The model \eqref{5.4} is known to be dual to $S^{(M)}_n [B ] $ of the type \eqref{3.1}, where 
$M =\sqrt{m^2 +e^2}$. The results of Section \ref{Section2} can naturally be extended to the case of the model \eqref{5.4}.

Supersymmetric extensions of \eqref{5.4} have been discussed in several publications
including \cite{DF,LS,K-tensor}. In particular, 
the massive tensor multiplet model \eqref{4.2} possesses the following generalisation:
\bea
S^{(m,e)}_{\rm tensor} [\J, \bar \J] = -\hf \int\rd^4x\rd^2\q\rd^2\qb\, E \, \big( G(\J)\big)^2 
- \hf m \Big\{ (m+\ri e)   \int\rd^4x\rd^2\q\, \cE \,\J^2
+{\rm c.c.} \Big\}~.~~~
\eea
Quantisation of this model can be carried out using the approach developed in section 
\ref{Section4}.

In conclusion, we would like to come back to the important work by Duff and van Nieuwenhuizen \cite{DvN}. Their argument concerning the quantum non-equivalence of the dual two-form and zero-form models in $d=4$  was based on the different trace anomalies. However, these theories are non-conformal and, therefore, the quantum operator $T^a{}_a$ 
``contains the effects of both classical and quantum breaking and is not equal to the trace anomaly'' \cite{GNSZ}. Nevertheless, the argument given in \cite{DvN} can be refined 
within a Weyl-invariant formulation for general gravity-matter systems
\cite{Deser,Zumino}.
We recall that a Weyl transformation acts on the covariant derivative as  
\bea
\nabla_a \to \nabla'_a =  \re^{\s} \Big( \nabla_a +\nabla^b\s M_{ba}\Big)~,
\eea
with the parameter $\s(x)$ being arbitrary.  
Such a transformation is induced by that of the gravitational field 
\bea
 e_a{}^m \to \re^\s e_a{}^m 
\quad \Longrightarrow \quad g_{mn} \to \re^{-2\s}g_{mn} ~.
\eea
In the Weyl-invariant formulation for gravity in $d\neq 2$ dimensions,
the gravitational field is described in terms of two gauge fields. 
One of them is the vielbein $e_m{}^a  (x)$ and the other is 
a conformal compensator $\vf (x)$. The latter is a nowhere vanishing scalar field 
with the Weyl transformation law 
\bea 
\vf \to \vf' = \re^{\hf (d-2)\s} \vf~. 
\eea
Any dynamical system is required to be invariant under these 
transformations.
 In particular, the Weyl-invariant extension of the Einstein-Hilbert gravity action is 
 \bea
S_{\rm GR}=\hf \int \rd^d x \,e \,\Big\{ \nabla^a \vf \nabla_a \vf +\frac{1}{4} \frac{d-2}{d-1} 
R \vf^2  
\Big\}~.
\label{WeylEH}
\eea
Choosing the Weyl gauge 
$
\vf = \frac{2}{\k}  \sqrt{\frac{d-1}{d-2}}
$
reduces \eqref{WeylEH} to the Einstein-Hilbert action.

The Weyl-invariant reformulation of the massless $p$-form action 
\eqref{massless-p-form} is
\bea
S_p [B;\vf  ] &=& -\frac{1}{2(p+1)! } \int \rd^d x \, e\, \vf^{2\D_p}
F^{a(p+1) } (B) F_{a (p+1)} (B)~, \quad \D_p =1-\frac{2p}{d-2}~.~~
\label{5.8}
\eea
Here we have made use of the Weyl transformation\footnote{There are two representations
for the $p$-form, 
$B_p = \frac{1}{p!} B_{m_1 \dots m_p} \rd x^{m_1} \wedge \dots \wedge 
x^{m_p}  =  \frac{1}{p!} B_{a_1 \dots a_p} e^{a_1}\wedge \dots \wedge 
e^{a_p} $, with $e^a =\rd  x^m e_m{}^a $ the vielbein. The $p$-form field with world indices, 
$B_{m(p)}$, is inert under the Weyl transformations.}
 of $B_p$
\bea
B'_{a(p) } = \re^{p\s} B_{a(p)}~.
\eea
Let $\G_p[\vf] $ be the effective action corresponding to 
\eqref{5.8}. Unlike the classical action \eqref{5.8}, 
the nonlocal functional $\G_p[\vf] $ is not Weyl invariant. However, this  Weyl anomaly
can be eliminated by adding a local  counterterm which  depends on the compensator 
$\vf$, see \cite{deWG} for the technical details.

At the classical level, the two massless theories  $S_p [B;\vf  ] $ and $S_{d-p-2} [A;\vf  ] $ prove to be dual, with $\D_p = - \D_{d-p-2}$.
In the even-dimensional case, 
it was shown in \cite{GKVZ} that the quantum effective actions 
for these theories, $\G_p[\vf] $ and $\G_{d-p-2}[\vf] $, 
are related to each other\footnote{The field $\f = - \D_p\ln   \vf  $ was interpreted in \cite{GKVZ} as the dilaton.}
 as
\bea
 \G_{d-p-2} [\vf] - \G_p [\vf] - (-1)^p {\mathfrak X}^{(m)} 
 \propto   \int \rd^d x \, e\,    \ln \vf \, \cE_d~,
 \label{5.10}
 \eea
where $\cE_d$ is the Euler invariant,
\bea
\cE_{d} = \frac{1}{(8\p)^n  n! } \ve^{a_1 b_1 \dots a_n b_n} 
\ve^{c_1 d_1 \dots c_n d_n} R_{a_1 b_1 c_1 d_1} \dots R_{a_nb_n c_n d_n} ~, 
\qquad d=2n~.
\eea 
Relation \eqref{5.10} is a generalisation of \eqref{B.8}. The expression on the right-hand side of \eqref{5.10} is a local functional and can be removed by adding a local counterterm. 
This proves the quantum equivalence of the theories.

In a similar manner supergravity in diverse dimensions can be formulated as 
conformal supergravity coupled to certain compensating supermultiplet(s) \cite{KakuT}.
The super-Weyl-invariant extensions of the
models \eqref{1.1} and \eqref{1.2} are given (see, e.g., \cite{CFG}) by
\begin{subequations}
\bea
S_{\rm tensor} [\J, \bar \J; S_0, \bar S_0] &=& -\hf \int\rd^4x\rd^2\q\rd^2\qb\, E \, 
\frac{\big( G(\J) \big)^2}{S_0 \bar S_0}  ~, \label{5.18a} \\
S_{\rm chiral} [\F, \bar \F; S_0, \bar S_0] &=& \hf \int\rd^4x\rd^2\q\rd^2\qb\, E \, 
S_0 \bar S_0(\F + \bar \F)^2~, \label{5.18b}
\eea
\end{subequations}
where $S_0$ is the chiral compensator, $\bar \cD_\ad S_0=0$, corresponding to the old minimal formulation for $\cN=1$ supergravity, 
see \cite{Siegel78,SG79,KU1,FGKV,KU2}.
By definition,  $S_0$ is nowhere vanishing and possesses the super-Weyl transformation $\d_\S S_0 = \S S_0$. The matter chiral scalar in \eqref{5.18b}
is super-Weyl neutral. The models \eqref{5.18a} and \eqref{5.18b} are classically equivalent. On general grounds, these models should be  equivalent at the quantum 
level. It would be interesting to carry out explicit calculations to check this. 
It should be pointed out that the actions \eqref{5.18a} and \eqref{5.18b}
lead to non-minimal operators for which the  standard superfield heat kernel 
techniques \cite{BK86,BK,McA} for computing effective actions do not work. 
Quantum supersymmetric theories with non-minimal operators were studied 
in \cite{BKS,K-20}.

Our analysis in this paper was restricted to those systems in  which the
classical  action is quadratic in the dynamical fields and therefore the corresponding effective action admits a closed-form  
expression in terms of the functional determinants of certain operators.  
In the case of nonlinear theories, such as the following model \cite{LR83,CFG} 
\bea
S &=&  \int\rd^4x\rd^2\q\rd^2\qb\, E \, \bigg\{ S_0 \bar S_0 \,
{\mathfrak F} \Big( \frac{G(\J)}{S_0\bar S_0} \Big)
+\frac{1}{16} V 
 \cD^\a (\bar \cD^2-4R) \cD_\a V
+ m  V  G(\J) \bigg\}
\eea
and its duals, it is not possible to obtain simple expressions for the  effective action. 
Nevertheless, the issue of quantum equivalence can still be addressed using the path integral 
considerations described by Fradkin and Tseytlin \cite{FT}.   
This approach was used in \cite{BK87} to prove quantum equivalence of the
Freedman-Townsend model \cite{FreedmanT} and the principal chiral $\sigma$-model.
\\


\noindent
{\bf Acknowledgements:}\\
We thank Fernando Quevedo for bringing important references to our attention. 
SMK is grateful to Ioseph Buchbinder and Jim Gates for email correspondence, and to Dmitri Sorokin for collaboration at an early stage of this project.
His work is supported in part by the Australian 
Research Council, project No. DP200101944.

\appendix

\section{Hodge-de Rham operator}\label{AppendixA}

Given a non-negative  integer $p \leq d$, 
the so-called Hodge-de Rham operator (also known as the covariant d'Alembertian)
\bea
\Box_p =- (\rd^\dagger \rd + \rd \rd^\dagger)
\eea
is defined to act on the space of $p$-forms. We recall that the operators of exterior derivative $\rd$ and co-derivative
$\rd^\dagger $ are defined to act  on a $p$-form $\o$ as 
\begin{subequations}
\bea
\rd : \quad \o_{a_1 \dots a_p} ~& \to & ~(\rd \o)_{a_1 \dots a_{p+1} } 
= (p+1) \nabla_{[a_1 } \o_{a_2 \dots a_{p+1}] } ~, \\
\rd^\dagger : \quad \o_{a_1 \dots a_p} ~& \to & ~
(\rd^\dagger \o)_{a_1 \dots a_{p-1} } = -\nabla^b \o_{b a_1 \dots a_{p-1}} ~.
\eea 
\end{subequations}
These operators are nilpotent, 
$\rd^2 =0$ and $(\rd^\dagger)^2 =0$, and are adjoint of each other, 
\bea
\frac{1}{(p+1)! } \int \rd^d x \, e\, 
(\rd \o )_{a_1 \dots a_{p+1} } \vf^{a_1 \dots a_{p+1} } 
= \frac{1}{p! } \int \rd^d x \, e\, 
\o_{a_1 \dots a_p} (\rd^\dagger \vf)^{a_1 \dots a_p } ~,
\eea
with respect to the inner product
\bea
\langle \o_p, \j_p \rangle 
= \frac{1}{p! } \int \rd^d x \, e\, 
\o_{a_1 \dots a_p} \j^{a_1 \dots a_p } ~.
\eea
In the case of a $d$-dimensional curved space $\cM^d$, the action of $\Box_p  $ 
 on a $p$-form $ \o_{a_1 \dots a_p} $ can be written as
\bea
\Box_p \o_{a_1 \dots a_p} = \nabla^b \nabla_b  \o_{a_1 \dots a_p} 
+\sum_{k=1}^{p} (-1)^k \big[ \nabla^b, \nabla_{a_k} \big]  
\o_{ba_1 \dots \widehat{a_k} \dots  a_p} ~.
\label{A.1}
\eea
The Hodge-de Rham operators  have the important properties
\bea
\rd \,\Box_p = \Box_{p+1} \rd~, \qquad \rd^\dagger\, \Box_p = \Box_{p-1} \rd^\dagger~,
\eea
which are used in Section \ref{Section2}.


\section{Massless $p$-forms in $d$ dimensions}\label{AppendixB}

Setting $m=0$ in \eqref{3.1} gives the massless $p$-form field theory
\bea
S_p [B ] &=& -\frac{1}{2(p+1)! } \int \rd^d x \, e\, 
F^{a_1 \dots a_{p+1} } (B) F_{a_1 \dots a_{p+1}} (B)~.
\label{massless-p-form}
\eea
The field strength $F_{p+1} (B) $ is invariant under gauge transformations 
\bea
\d_\z B_{a(p)} = p \nabla_{[a_1} \z_{a_2 \dots a_p]} ~,
\eea
and so is the action. It is known, by Poincar\'e duality, that the massless gauge theories with actions $S_p [B ] $ and $S_{d-p-2} [B ] $ are classically equivalent.  

The energy-momentum tensor, $T^{ab}_p (B)$,  
corresponding to \eqref{massless-p-form}
is obtained from \eqref{EMtensor} by setting $m=0$. 
\bea
T^{ab}_{p} (B)&=& \frac{1}{p!} \Big\{ F^{a c_1 \dots c_p} (B) F^b{}_{c_1 \dots c_p} (B)
-\frac{1} {2(p+1)} \eta^{ab}  F^{c_1 \dots c_{p+1}} (B) F_{c_1 \dots c_{p+1}} (B) \Big\} 
~.
\label{EMtensor-massless}
\eea
It is conserved on-shell, 
$\nabla_b T^{ab}_{p} =0$.
If the dimension of spacetime is even $p+1 = d/2$, 
the energy-momentum tensor is traceless in the massless case, 
\bea
d=2(p+1) \quad \implies \quad \eta_{ab} T^{ab}_{d/2 -1} =0~.
\eea
This is a corollary of the fact that the $p$-form action in $2(p+1)$ spacetime dimensions 
\bea
S_p [B ] &=& -\frac{1}{2(p+1)! } \int \rd^{2(p+1)} x \, e\, 
F^{a_1 \dots a_{p+1} } (B) F_{a_1 \dots a_{p+1}} (B)
\eea
is invariant under arbitrary Weyl rescaling of the vielbein.

Let $\G_p$ denote the effective action for the massless $p$-form theory 
\eqref{massless-p-form}. As shown in \cite{Obukhov} (see also \cite{Siegel}), 
\bea
\G_p = \frac{\ri}{2} \sum_{k=0}^{p} (-1)^k (1+k) \,\ln \det \Box_{p-k}  ~.
\label{B.1}
\eea 
In the case that $p =d-1$, the action describes no local degrees of freedom, and therefore 
the corresponding effective action  should be a topological invariant. Indeed, making use of \eqref{B.1} allows us to
rewrite $\G_{d-1}$ in the form 
\bea
\G_{d-1} = -\frac{d}{2} {\mathfrak X}~, \qquad
{\mathfrak X} = \frac{\ri}{2} \sum_{k=0}^d (-1)^k
\ln \det \Box_k ~.
\label{B.2}
\eea
The functional $\mathfrak X$ is obtained from ${\mathfrak X}^{(m)}$ given by 
eq. \eqref{TI1}
by setting $m=0$. 

For $p \neq d-1, d$, it is known that the massless  $p$-form and $(d-p-2)$-form 
models 
are classically equivalent. For the corresponding effective actions, 
the following relation holds:
\bea
 \G_{d-p-2} - \G_p=  (-1)^p \Big( \frac{d}{2} - p -1\Big) {\mathfrak X}~.
 \label{B.8}
\eea
This result was established in  \cite{FT,GNSZ,BK88} for $d=4$, and later generalised to 
the $d>4$ case in \cite{GKVZ,Barbon}.


\section{Massless three-form in four dimensions} \label{AppendixC}

The analysis described in the previous appendix has some nuances in the $d=p+1$ case. For simplicity, here we discuss the massless three-form in four dimensions.

The gauge three-form model is described by the action 
\bea
\widetilde{S}_3[V] = \hf \int_{\cM} \rd^4 x \, e\, H^2 -  \int_{\cM} \rd^4 x \, e\, \nabla_a (V^a H) \equiv S_3[V] -  \int_{\cM} \rd^4 x \, e\, \nabla_a (V^a H) 
~,
\eea
where $H:= \nabla_a V^a$ is the field strength being invariant under gauge transformations
\bea
\d_B V^a = \hf \ve^{abcd} \nabla_b B_{cd} ~.
\eea
The second term in the action is a boundary term; it was introduced in\cite{DJ,GLS}. To obtain a consistent variation problem, one demands \cite{DJ} that 
\bea
\d H \Big|_{\pa \cM} =0~,
\eea
such that an arbitrary variation of the action is 
\bea
\d \widetilde S_3[V] = - \int_{\cM} \rd^4 x \, e\, \d V^a \nabla_a H 
- \int_{\cM} \rd^4 x \, e\, \nabla_a( V^a \d H) ~.
\eea
The equation of motion is
\bea
\nabla_a H =0~ \quad \implies \quad H = c = {\rm const}~.
\eea
This shows that the model under consideration has no local degrees of freedom.

Different values of $c$ correspond to different vacua in the quantum theory. When computing the path integral, for a given $c$ we make use of the background-quantum splitting 
\bea
V^a = V_0^a +v^a~, \qquad \nabla_a V^a_0 =c~,
\eea
such that the classical action becomes
\bea
\widetilde{S}_3[V] = - \hf c^2 \int_{\cM} \rd^4 x \, e + S_3[v] ~.
\eea
Here the first contribution on the right is the cosmological term. 
Evaluating the path integral, for the effective action one gets
\bea
\G_3 [g_{mn}] =  - \hf c^2 \int_{\cM} \rd^4 x \, e -2 \mathfrak{X}~, 
\label{2.8}
\eea
where we have defined 
\bea
\mathfrak{X} := \frac{\ri}{2} \ln \frac{  \det \Box_2\, [\det \Box_0]^2}{ [\det \Box_1]^2 }
~.
\label{TopInv}
\eea
The functional $\mathfrak X$ is the four-dimensional version of the 
topological invariant \eqref{B.2}.


\section{Duality  with topological mass term} \label{AppendixD}

To construct a dual formulation for \eqref{5.1}, we introduce the first-order action
\bea
S [L_q, A_{q} , C_{q-1}  ] &=& \frac{1}{q!}  \int \rd^d x \, e\, \bigg\{
\hf L^{a (q) }  L_{a (q)}
-\frac{1}{2(q+1)} F^{a(q+1) } (A) F_{a (q+1)} (A) \non \\
&& + L^{a(q)} \Big[ m A_{a(q)} + F_{a(q)} (C) \Big]  \bigg\}~,
\label{D.1}
\eea
where $L_{a(q)} $ and $C_{a(q-1)}$ are unconstrained antisymmetric tensor fields.
The equation of motion for  $C_{a(q-1)} $ implies that 
$L^{a(q)} = \frac{1}{(p+1)!} \ve^{ a(q) b(p+1)} F_{b(p+1) } (B) $, and then 
the action \eqref{D.1} turns into \eqref{5.1}. On the other hand, we can eliminate 
$L_{a(q)} $ from the action \eqref{D.1} using the corresponding equation of motion.
This leads to 
\bea
S^{(m)}_q [A_{q} , C_{q-1}  ] &=& -\frac{1}{2 q!}  \int \rd^d x \, e\, \bigg\{
\frac{1}{q+1} F^{a(q+1) } (A) F_{a (q+1)} (A) 
\non \\
&& 
\qquad + \Big[ m A_{a(q)} + F_{a(q)} (C) \Big]^2  \bigg\}~,
\label{D.2}
\eea
This is the Stueckelberg formulation for the massive $(d-p-1)$-form model, see eq. 
\eqref{3.7}. Thus we have shown that the massive $q$-form model \eqref{D.2} is dual to \eqref{5.1}.

There is an alternative dual formulation for \eqref{5.1}, which is obtained 
by making use of the first-order action
\bea
S [L_p, B_{p} , V_{p-1}  ] &=& \frac{1}{p!}  \int \rd^d x \, e\, \bigg\{
\hf L^{a (p) }  L_{a (p)}
-\frac{1}{2(p+1)} F^{a(p+1) } (B) F_{a (p+1)} (B) \non \\
&& + (-1)^{ d+ dp} L^{a(p)} \Big[ m B_{a(p)} + F_{a(p)} (V) \Big]  \bigg\}~,
\label{D.3}
\eea
where $L_{a(p)} $ and $V_{a(p-1)}$ are unconstrained antisymmetric tensor fields.
The equation of motion for  $V_{a(p-1)} $ implies that 
$L^{a(p)} = \frac{1}{(q+1)!} \ve^{ a(p) b(q+1)} F_{b(q+1) } (A) $, and then 
the action \eqref{D.3} turns into \eqref{5.1}.
On the other hand, integrating out
$L_{a(p)} $ leads to the the massive $p$-form model 
\eqref{3.7}. 


\begin{footnotesize}

\end{footnotesize}


\end{document}